**Title of the paper: A new DFM approach to combine machining and additive manufacturing**

**Authors: Olivier Kerbrat, Pascal Mognol, Jean-Yves Hascoët**

**IRCCyN (Institut de Recherche en Communications et Cybernétique de Nantes)**

1 rue de la Noë – BP 92101 – F-44321 Nantes Cedex 03, Nantes, France

Corresponding author: Olivier Kerbrat

IRCCyN / MO2P team

1 rue de la Noë

BP 92101

F-44321 Nantes Cedex 03

France

Tel.: +33(0)299055275

Fax: +33(0)299059328

E-mail address: Olivier.Kerbrat@irccyn.ec-nantes.fr

# A new DFM approach to combine machining and additive manufacturing

## 1. Introduction

In order to stay competitive in the modern mass production industry, products have to be designed and manufactured with the following opposing goals: decreasing time and cost; improving quality and flexibility. One way to improve product competitiveness is the Design For Manufacturing (DFM) approach. DFM involves simultaneously considering design goals and manufacturing constraints in order to identify manufacturing problems while parts are being designed; thereby reducing the lead time for product development and improving product quality [1,2]. Most of the DFM systems do not have the ability to handle multiple processes, and concentrate only on one specific manufacturing process. This paper aims to bring a new DFM approach to multi-process manufacturing. Nowadays, technical improvements in additive manufacturing processes provide the opportunity to manufacture real functional metal parts [3-6]. In fact, rapid manufacturing supersedes rapid prototyping because the additive technology, such as Selective Laser Sintering (SLS), is no longer exclusively used for prototyping. Furthermore, new fabrication possibilities are offered by machines that are able to depose and fuse metal powder directly on machined blocks. These additive manufacturing processes provide an interesting alternative to High-Speed Machining (HSM). Difficult – or even impossible – to machine parts (such as very complex shapes and conformal cooling channels in injection dies) may be manufactured by an additive process in preference to a costly and time consuming electro-discharge machining process. The problem is the characterization of the manufacturing complexity at the design stage in order to choose the most appropriate process.

From our vision, parts could advantageously be designed with modular and hybrid points of view in which parts are seen as 3-D puzzles with modules realized separately and further assembled. This hybrid modular concept [7] has several advantages:
- All the modules may be produced simultaneously and independently;
- Several alternatives of the same product may be easily manufactured changing only one module to provide new part functions instead of the whole product;
- Each module of the part is realized by the best manufacturing process, in term of time, cost and/or quality. In fact, additive manufacturing processes (such as SLS or powder projection) have to be compared to HSM process in order to choose the best way to obtain each module;
- Manufacturing difficulties are reduced because some small modules may be easier to manufacture than a complex one.

The two main drawbacks of the hybrid modular concept are:
- All the modules must be carefully gathered in order to create a whole part with the same level of quality as a one-piece part;
- The choice of a hybrid modular design instead of the traditional one-piece one is still not easy to do at the design stage because no DFM system is able to bring qualitative information on manufacturing complexity for different processes.

The first drawback has been previously studied with the definition of standard assemblies for multi-component prototypes [7], and this paper concentrates on the second one.

The aim of this work is to propose a new DFM approach, combining additive processes (such as SLS or powder projection) to more traditional subtractive ones (HSM) in a hybrid modular vision. A hybrid modular design methodology is created, integrating manufacturability issues at the design stage. This paper explains what this methodology is based on (section 2), how it can be used in CAD software

(section 3) and tests have been carried out on two industrial parts, taken from the field of tooling (section 4).

**2. Hybrid modular design methodology**

Two points have to be taken into account in the creation of the hybrid modular design methodology: a manufacturability evaluation and a hybrid modular optimization that can improve the manufacturability. The first point is detailed in the two first sub-sections; it concerns the calculation of manufacturability indexes. The second point is detailed in the third sub-section. A schematic view of the methodology is then explained in part four.

*2.1. Manufacturability evaluation*

In traditional DFM approaches, there are many different scales on which manufacturability can be measured: binary, qualitative, quantitative and ratings based on manufacturing time and cost. The most basic scale is a binary one: it simply reports whether or not a given set of design attributes is feasible. For example, the system underlines whether a manufacturing process might be applicable for the features modeling a part [9]. On a qualitative scale, designs are given grades based on their manufacturability by a certain process. For example, "good", "bad", "marginal" may be used as design ratings [10]. Such evaluations are hard to interpret and difficult to combine in the case of different parts or different manufacturing processes. The quantitative scale assigns a numerical value, for example between 0 and 1 [11]. Even it can be difficult to interpret such measures, this scale allows comparisons between several alternative designs of similar products. Another way of estimating manufacturability is to associate ratings based on manufacturing time and cost. They present a realistic view of the difficulties in manufacturing a proposed design and can be used by the designer to help him in designing products that meet target production time and cost [1]. Nevertheless, the evaluations of manufacturing time and cost are major issues at the design stage. In fact, it has been proven that machining cycle time predictions given by CAM software are inaccurate because CAM systems ignore the dynamics of the machine tool; therefore they are unable to estimate the actual cycle time at high feed rates for complex shapes [12-14]. In the same way, in design activities, the lack of information about the cost structure and the production process plans do not help designers to make precise cost estimations [15].

Consequently, for the manufacturability evaluation of the hybrid modular design method, a quantitative measure of manufacturing difficulties, for both additive and subtractive processes, has to be carried out. In this paper, manufacturability evaluation is based on the calculation of manufacturability indexes; they are derived from design parameters which have a great influence on manufacturing difficulties. One essential point is to take into account all of the design parameters (geometry, dimensions, material information, specifications) which can be determined only with CAD model. In fact, parameters which require a complete manufacturing preparation analysis (for example: cutting-tool path strategy) are not taken into account, so as not to depend on manufacturer's skills.

*2.2. Manufacturability indexes*

The manufacturability indexes, calculated from the design parameters, are classified into two categories: global / local. As an example, a manufacturability index may be calculated from the parameter "Volume". In fact, volume has a great impact on production time in an additive manufacturing process. That is why a bigger part will be considered more difficult to manufacture than a smaller one, in case of an additive process, because manufacturing time will take longer.

This global view of manufacturability cannot satisfy the complete analysis of manufacturing difficulties, because often there are few part's details that can change the choice of the manufacturing process (a curve radius of a small complex shape, for instance). This consideration forces the introduction of local indexes, which are defined for each area of the part. The CAD model has to be decomposed into several elementary elements, and local indexes are calculated for each element. The most common method to decompose a CAD model into basic components is certainly the feature decomposition. Most of the studies on DFM methods imply using a feature decomposition of the part CAD model, then associating manufacturability evaluation with each feature. The major problem is that features usually rely on one specific field. As an example, machining features are developed for CNC machining [16], but manufacturing features for additive technologies are still under development [17]. Furthermore, for free-form surfaces, features do not bring enough information to the shape. So the local manufacturability indexes cannot be based on feature decomposition.

The decomposition accuracy must be of a high level for the areas that are geometrically complex (with many changes in surface orientations), whereas it may be lower for quite simple areas (a plane section for instance). Octree decomposition [18] is a good candidate for part CAD model decomposition. An octree is a tree data structure, which represents a three-dimensional object by the recursive division of space into 8 small cubic cells or small parallelepipeds, named octants. The octants are classified into three categories: black, white and grey. Black octants are those that are completely included in the object of interest whereas white ones are those that are completely exterior to the object. Grey octants are those that are partially inside and outside the object. The subdivision process is performed on grey octants until a desired resolution is reached.

The main advantages are:
- The size of each cell depends on the local geometric complexity of the object represented;
- This decomposition is neutral (it neither depends on a specific manufacturing process nor on a specific CAD software);
- Decomposition models can acquire high accuracy relatively quickly.

An example of octree decomposition is shown in Fig. 1.

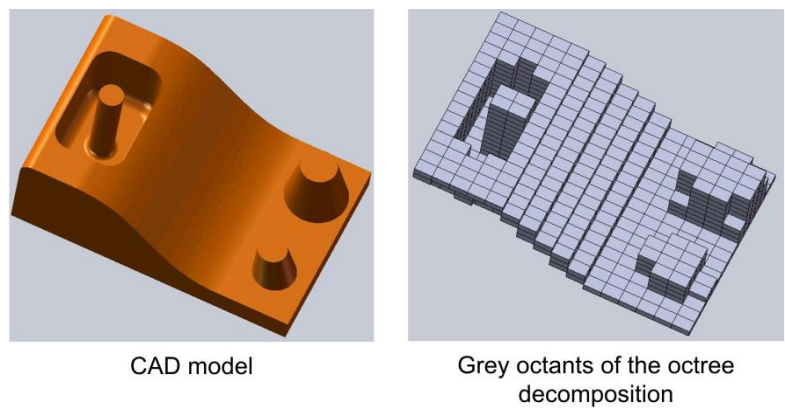

**Fig. 1.** A CAD model and its octree decomposition (only grey octants are shown).

Local indexes will be calculated for each grey octant which makes up the octree decomposition of the part CAD model. Then a color map of manufacturability is obtained by coloring the fraction of the part contained in each grey octant according to an appropriate color scale. This map provides an accurate view of the most difficult to manufacture areas of the part. The blue areas correspond to the easiest to manufacture areas (the lowest value of local manufacturability index) whereas the red ones correspond to the most difficult to manufacture ones (the highest value). An example is given in Fig. 2, which is a map of manufacturing difficulties from a local manufacturability index based on the flexibility of the cutting-tool when considering a 3-axis machining process. In order to facilitate the visualization, black octants are also plotted in blue.

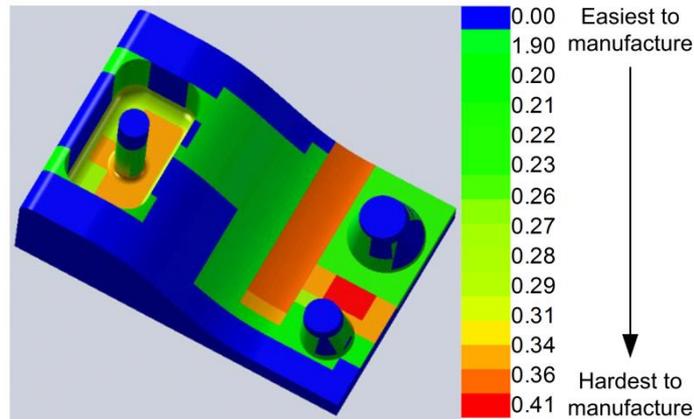

**Fig. 2.** Map of manufacturing difficulties and the associated color scale.

Table 1 sums up the global and local indexes that have been developed in our DFM system, for additive and subtractive processes. The indexes are based on an analysis of which design parameters have a great influence on time, cost and quality in case of a machining or an additive manufacturing process. In this paper, the methods used to calculate the values of these indexes are not detailed, but an overall explanation and the development of algorithms can be found in [19]. These indexes have values between 0 and 1, without dimension. They provide information to decide which parts or which areas of a part are the most complex to manufacture by one specific process. Other indexes may be easily added in this table, in particular those which are based on surface orientations, material information for additive manufacturing and specifications (dimensional, geometric and location tolerances), but they are not discussed here.

| Category | Index | Linked to | Type | Process |
|---|---|---|---|---|
| Geometric parameters | $C(d)^-$ | Maximal dimension | Global | Machining |
| | $C(c)^-$ | Quantity of chips | Global | |
| | $C(f)^-$ | Cutting-tool flexibility | Local | |
| | $C(d)^+$ | Maximal dimension | Global | Additive manufacturing |
| | $C(v)^+$ | Volume | Global | |
| | $C(s)^+$ | Skin surface | Global | |
| | $C(h)^+$ | Height | Local | |
| | $C(\rho)^+$ | Distance from the platform centre | Local | |
| Material information | $C(m)^-$ | Material hardness | Global | Machining |
| Specifications | $C(r)^-$ | Surface roughness | Global | Machining |

**Table 1.** Manufacturability indexes.

The values of the manufacturability indexes (global and local) and the maps of manufacturing difficulties provide a well-detailed view of the manufacturability, seen directly from a CAD model. They are calculated with a new manufacturability analysis system developed in CAD software.

*2.3. Hybrid and modular approaches to improve the manufacturability*

According to the manufacturability analysis results, two possibilities may be used to reduce manufacturing difficulties:
- A modular approach, with modules realized aside and further assembled;
- A hybrid approach, with different manufacturing processes chosen for the different zones of the part;

Of course, these possibilities must be considered simultaneously. The choice depends on which parameters are involved in the most difficult to manufacture areas. For example, if a high value for the $C(f)^-$ index (local index based on cutting-tool flexibility) comes from a low curve radius value for a concave shape, a modular approach will not reduce manufacturing difficulties in this particular zone of the part, but a hybrid point of view, considering an additive manufacturing process for the areas with low radius values may improve manufacturability. Fig. 3 shows an academic example of this hybrid approach. A map of manufacturing difficulties was drawn up with the DFM system, and some areas seemed to be impossible to machine (the red ones). Consequently, a hybrid design was proposed. One module was machined and another one keept the previous impossible to machine zones and was realized by an additive fabrication process (SLS for instance).

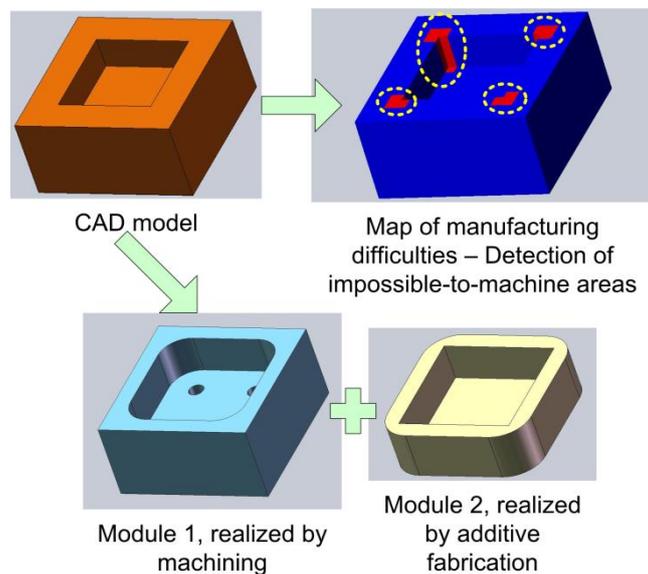

**Fig. 3.** Academic example of the hybrid approach in reducing manufacturing difficulties.

On the other hand, creating modules, realized individually, using the same process and then assembled, may also decrease manufacturing difficulties in some particular cases. For example, if the part is larger than the fabricating volume of the additive fabrication machine (in this case the index $C(d)^+$ will be equal to 1), then a modular point of view would help in creating a modular part CAD model with modules compatible with the machine. Another academic example is shown in fig. 4, where the manufacturability is affected by the large amount of chips (represented by the index $C(c)^-$) A modular design will then provide a significant improvement in manufacturability. In fact, machining the two modules separately will generate fewer chips than machining the part in one piece, consequently manufacturing complexity is considered to be reduced.

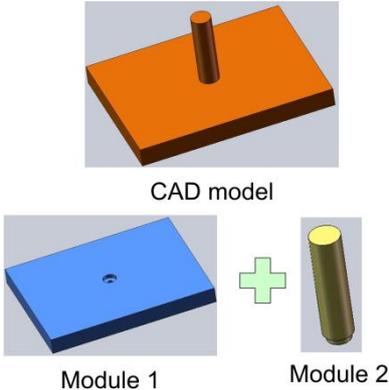

**Fig. 4.** Academic example of the modular approach for improving manufacturability.

A third possibility for improving manufacturability is a modification in design parameters. For example, if the difficulties result from a low curve radius value for a concave shape (as in Fig. 3), the designer might change the value of the parameter "radius" in order to reduce manufacturing difficulties and check if its design fits the requirements. This possibility is no longer studied in this paper because the focus is put on hybrid and modular points of view.

*2.4. Schematic view of the methodology*

A DFM methodology is then created, in which the analyses of manufacturability are achieved and both points of view (modular and hybrid) help in decreasing the manufacturing difficulties.
This method is divided into 6 stages, a schematic view is proposed in Fig. 5.

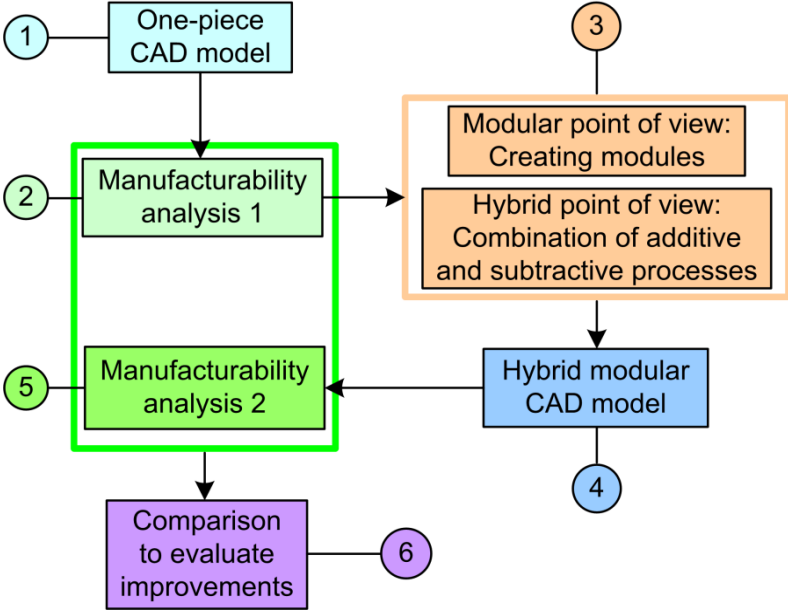

**Fig. 5.** Hybrid modular design methodology.

The starting point of this methodology is the one-piece CAD model. The manufacturability of this one-piece part is evaluated directly in CAD software with the help of the manufacturability indexes calculation. Then the modular and hybrid points of view are taken into account in order to create a hybrid modular part. The manufacturability analysis is performed on this new CAD model. The last stage is a comparison between the two manufacturability analyses in order to quantify the advantages of the hybrid modular design.

## 3. Development in CAD software

This methodology has been implemented in CAD software (SolidWorks by Dassault Systemes) with Visual Basic language. The procedure shows the results of the manufacturability analyses (in stage 2 and 5) and the comparison between alternative designs (stage 6). The results are exported as Microsoft Excel files in order to be easily read by designers.

For global indexes, the system directly posts the values. The higher the value, the more difficult it is to manufacture the part. In the case of a local index, one step of octree decomposition is carried out. Then the index values are calculated for all the grey octants of the octree decomposition. For each octant, the higher the index value, the more difficult it is to machine the fraction of the part contained in the octant. In order to represent the manufacturability distribution, a map of manufacturing difficulties is displayed in CAD software (with automatic or customized color scale). If the accuracy of the decomposition is not satisfactory, another level of octree decomposition is carried out. The index values are again calculated for new octants, and a more detailed map of manufacturing difficulties is obtained.

For each local index $C(i)$, two global values are also calculated: the maximal value $C(i)_{max}$ and a mean value $C(i)_{mean}$ (Equation 1):

$$C(i)_{mean} = \frac{\sum_j (C(i)_j \times V_j)}{\sum_j V_j} \quad \text{(Equation 1)}$$

where $V_j$ is the volume of the fraction of the part contained in the octant for which $C(i)_j$ is being calculated.

A picture of the interface of the system is given in Fig. 6. It corresponds to the manufacturability analysis for additive manufacturing. With this interface, the user may select which index he wants to calculate.

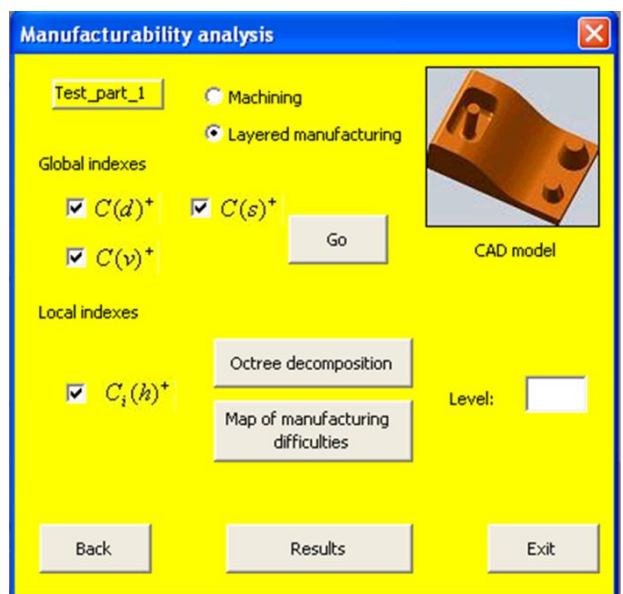

**Fig. 6.** Calculation system of manufacturability indexes.

## 4. Results and discussion: examples of industrial parts in the field of tooling

In order to validate this new DFM system, two industrial parts have been studied. These examples have been chosen in the field of tooling (dies and moulds). The reason is that even if machining is still the most common process used to manufacture dies and moulds, additive fabrication provides an interesting alternative method. In fact, complex metal parts (such as dies and moulds) represent good candidates for additive manufacturing. That is why the 6 stages of the hybrid modular design methodology are applied on two industrial tools and the results discussed.

*4.1. Industrial modular injection mould*

4.1.1. One-piece CAD model (stage 1)

The first example concerns a part of an industrial injection mould: a core for investment casting. Fig. 7 presents the one-piece CAD model, corresponding to the initial step of the methodology.

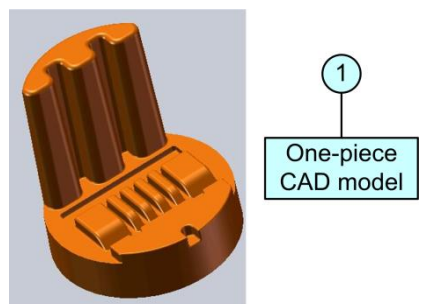

**Fig. 7.** One-piece CAD model.

4.1.2. Manufacturability analysis for machining (stage 2)

The designers of this core have identified difficulties in machining this part, caused by low rigidity in cutting-tool (due to the geometry, the cutting-tool that is required to machine this part must be very long). Instead of choosing another manufacturing process, such as electro-discharge machining for example, a modular point of view will help designers to create the part in two modules which will be machined aside and further gathered. So the hybrid modular tool design methodology is used. A manufacturability analysis is carried out using the procedure implemented in CAD software and the results are presented in Fig. 8.

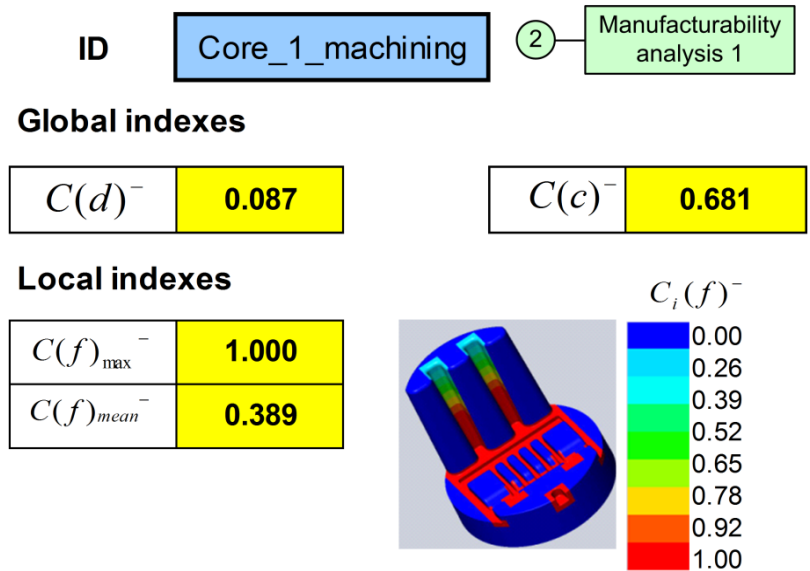

**Fig. 8.** Manufacturability analysis for this part.

4.1.3. Modular point of view (stages 3 and 4)

The results provide information on how difficult it is to manufacture this part by machining, and how manufacturability can be improved. Firstly, the global index $C(d)^-$ has a low value (because the tool is quite small), consequently it is not interesting to consider a decrease for this index. Next, concerning the global index $C(c)^-$, it can be noticed that the value tends to be relatively high so it may be interesting to look for a better design so as to reduce the quantity of chips (a modular approach will bring a drop in manufacturing difficulties for this particular index). Then, focus is put on the local index $C(f)^-$. The maximal value is 1; it implies that there are areas of the part that will be very difficult to machine and even impossible to machine. Furthermore, the map of the manufacturing difficulties (Fig. 7) shows where these specific areas are and what has to be done to improve manufacturability in these particular zones of the part. This consideration provides a modular design of the core which is presented in Fig. 9. In this example, the assembly process is not treated because it is not within the range of this paper. It would be necessary to take into account the constraints of assembly techniques between modules, and for instance, standard assemblies should be automatically defined in stage 4 of the methodology as it can be found in [8]

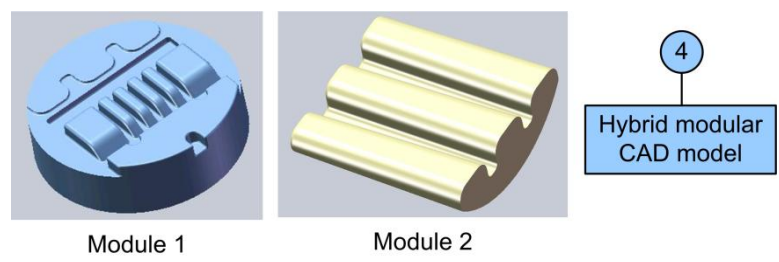

**Fig. 9.** Modular design for this core.

4.1.4. Manufacturability analyses for the modules (stage 5)

The fifth stage of the methodology is constituted by the manufacturability analyses of the two modules. Results are given in Fig. 10.

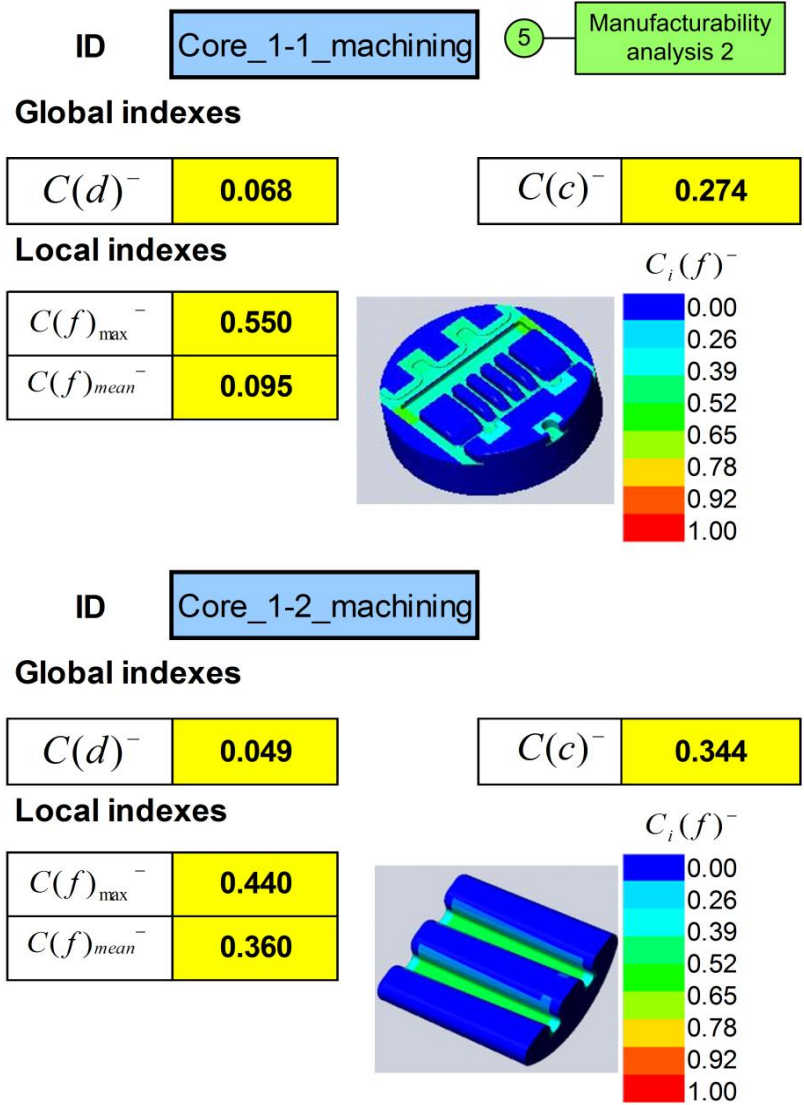

**Fig. 10.** Manufacturability analyses for the two modules.

In order to compare the modular design to the one-piece one, modules have to be gathered. The values of each index $C(i)$ for all the modules are brought together in a total value $C_{\text{total}}(i)$ with Equation 2:

$$C_{\text{total}}(i) = \sum_{j=1}^{N_{\text{mod}}} \omega_j C(i)_j \qquad \text{Equation 2}$$

Where $N_{\text{mod}}$ represents the number of modules composing the part (in this example, $N_{\text{mod}} = 2$), $\omega_j$ corresponds to the weight associated to module $j$ and $C(i)_j$ is the value of index $C(i)$ for module $j$. For a first approach, weights $\omega_j$ are determined with Equation 3:

$$\forall j \in \{1,\ldots, N_{\text{mod}}\}, \omega_j = \frac{V_j}{\sum_{j=1}^{N_{\text{mod}}} V_j} \qquad \text{Equation 3}$$

The calculations of the total values are done for global indexes and mean values of local ones. In case of maximal values of a local index, $C_{\text{total}}(i)$ corresponds to the highest maximal value of all the modules.

Table 2 shows the results for this two-module industrial core.

|  | **Core_1** | | |
|---|---|---|---|
| **Index** | **Core_1-1** | **Core_1-2** | **Total value** |
|  | $\omega_1 = 0.67$ | $\omega_2 = 0.33$ |  |
| $C(d)^-$ | 0.068 | 0.049 | **0.062** |
| $C(c)^-$ | 0.274 | 0.344 | **0.297** |
| $C(f)_{max}^-$ | 0.550 | 0.440 | **0.550** |
| $C(f)_{mean}^-$ | 0.095 | 0.360 | **0.182** |

**Table 2.** Total values for manufacturability indexes.

4.1.5. Comparison between the one-piece design and the two-modules one (stage 6)

The last stage of the methodology is constituted by the comparison of the analyses obtained in stage 2 and in stage 5 in order to evaluate quantitatively the improvements provided by the modular approach. The comparison is done directly between the values of the manufacturability indexes calculated for the first one-piece CAD model and the total values for the modules of the modular CAD model.
The results are presented in Fig. 11.

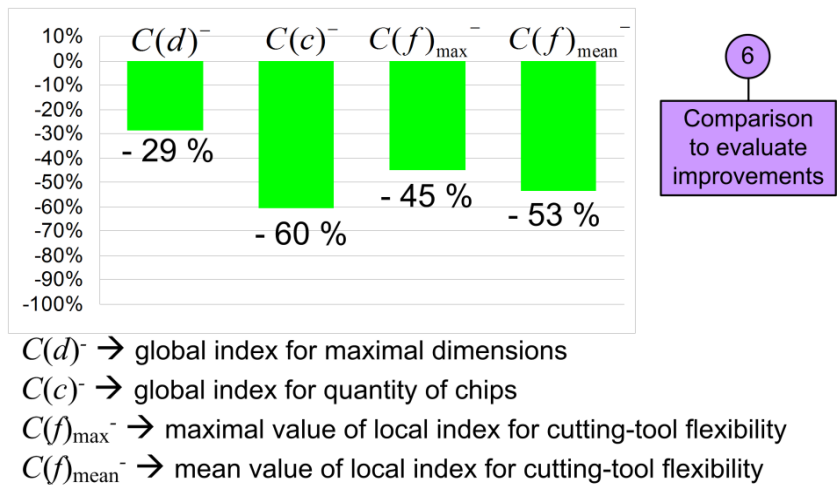

$C(d)^-$ → global index for maximal dimensions
$C(c)^-$ → global index for quantity of chips
$C(f)_{max}^-$ → maximal value of local index for cutting-tool flexibility
$C(f)_{mean}^-$ → mean value of local index for cutting-tool flexibility

**Fig. 11.** Comparison between the two analyses.

This picture shows the evolutions in the indexes, it can be seen that all the indexes values are reduced. Moreover, the maps of manufacturing difficulties give evidence of the significant decreases in manufacturing difficulties in the previous most difficult to machine areas. The modular design has considerably reduced the quantity of chips (- 60 % in $C(c)^-$), and allows the use of less flexible cutting-tools, which can be observed by a decrease in $C(f)_{max}^-$ (- 45 %) and $C(f)_{mean}^-$ (-53 %).
These decreases in manufacturing difficulties will have something of an impact in time, cost and quality of the tool. This impact is not quantified because it is tricky at the design stage and the assembly constraints are not taken into account in this example, so only the manufacturing difficulties evolutions can be quantified. That is why the methodology merely takes an interest in manufacturability, and not directly in time, cost and quality. In this example, the evolutions in the indexes are enough significant to conclude that the methodology provides an interesting design alternative with the modular approach, even if assembly constraints are not quantified.

*4.2. Industrial automotive die*

4.2.1. Presentation of the die

Another way of using this system is for a comparison of two manufacturing processes (additive and subtractive) for one CAD model in order to determine which zones of a part may advantageously be machined or realized by an additive manufacturing process. This possibility is illustrated by an industrial example taken from automotive industry. The part, presented in Fig. 12, has the following dimensions: 630 x 182 x 100 mm. It is a stamping die for producing sheet metal parts for motor vehicles.

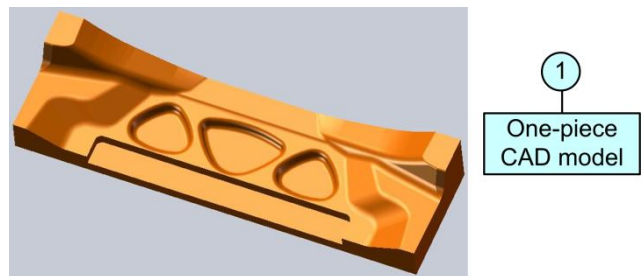

**Fig. 12.** One-piece CAD model.

4.2.2. Manufacturability analyses

Manufacturability indexes are calculated with the manufacturability analysis system for machining (Fig. 13).

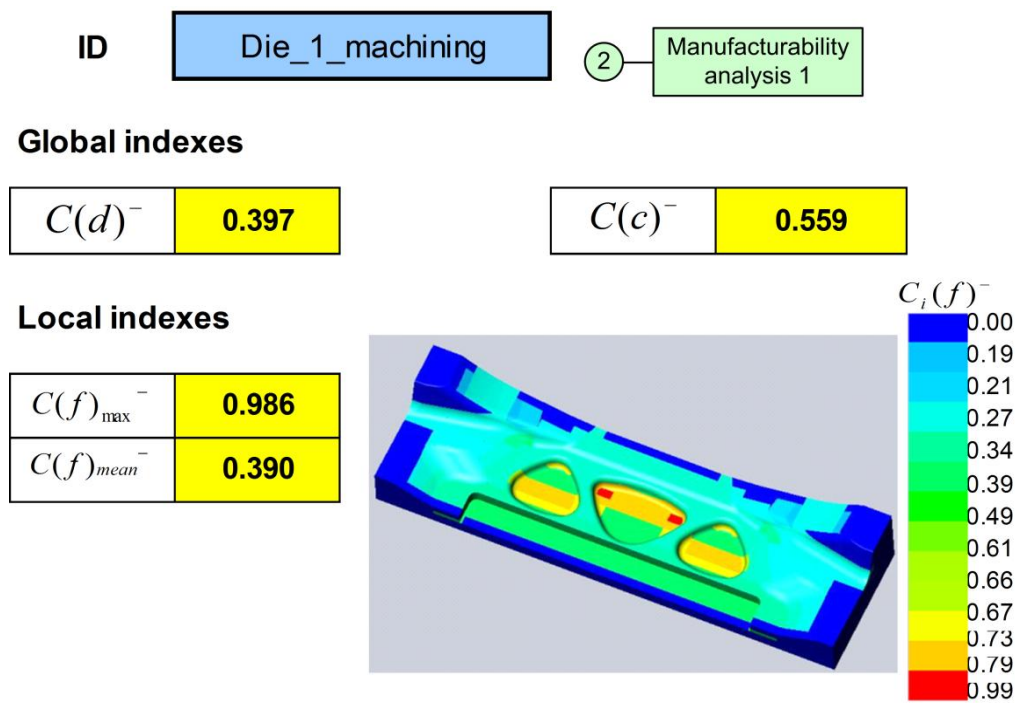

**Fig. 13.** Manufacturability analysis for machining.

In this second example, it can be seen that some areas of the die are very difficult to machine, according to the flexibility index high value, at the corner radii of the three pockets ($C(f)_{max}^- = 0.986$, really closed to its maximal value, 1). For these areas, a modular approach, similar to the previous example, will not bring a significant decrease in index values. In order to reduce the manufacturing difficulties, an additive process could be used for this die realization. So a second manufacturability analysis is done, the values of the manufacturability indexes for additive fabrication are calculated. The results are presented in Fig. 14.

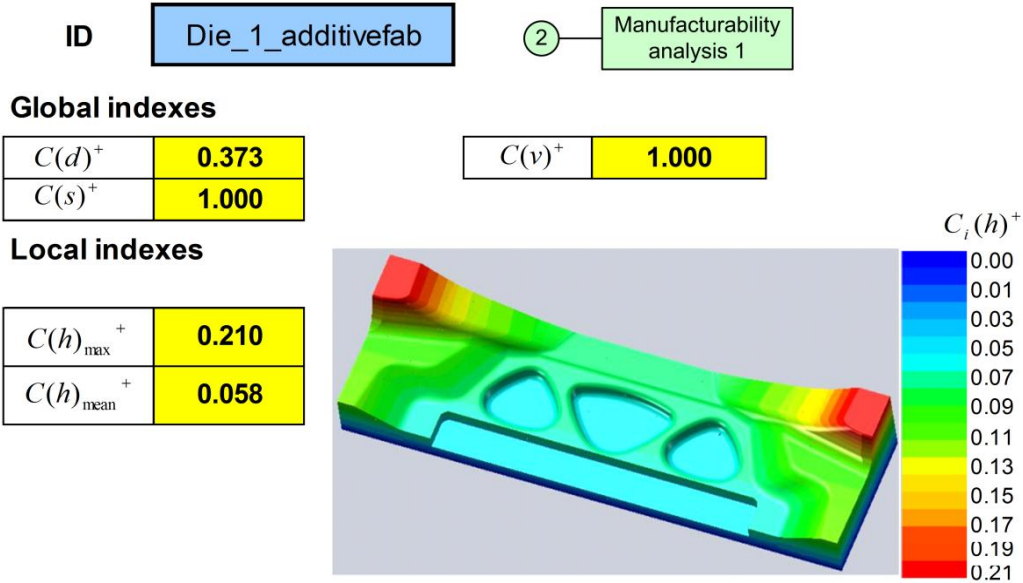

**Fig. 14.** Manufacturability analysis for additive manufacturing.

4.2.3. Hybrid point of view

The map of manufacturing difficulties corresponding to the local index $C(h)^+$ (linked to the height of the tool) shows that some areas of this part are quite difficult to manufacture by an additive process. Furthermore, the global indexes have high values mainly because of the large dimensions of the part, which imply using a costly machine with a very large building volume. However, it can be seen on the map that the previous most difficult to machine zones are quite easy to manufacture by an additive process. These analyses indicate that the two manufacturing processes (subtractive and additive) have to be combined in order to produce a hybrid modular die with an improved manufacturability. Consequently, the areas which are the most difficult to machine will advantageously be realized by an additive fabrication technology, and the areas that are easy to machine will be realized by a machining process, as it is presented in Fig. 15. It can be noticed that the progress in additive fabrication allow to consider a metal deposition directly on a machined support. Thus module 1 will then be machined and module 2 corresponds to the part that will be realized by powder projection directly on module 1.

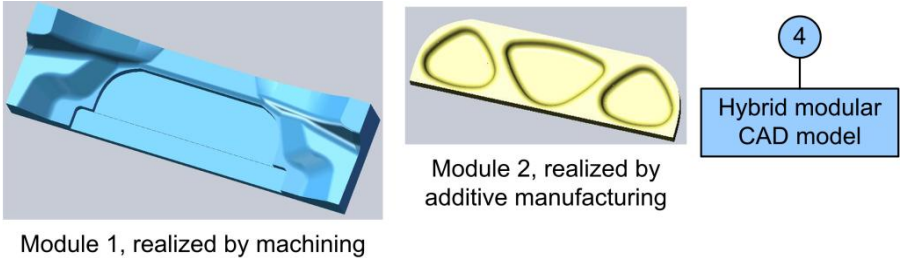

**Fig. 15.** Hybrid modular CAD model.

The next stage of the methodology is the manufacturability analysis of the modules. The results are presented in Fig. 16.

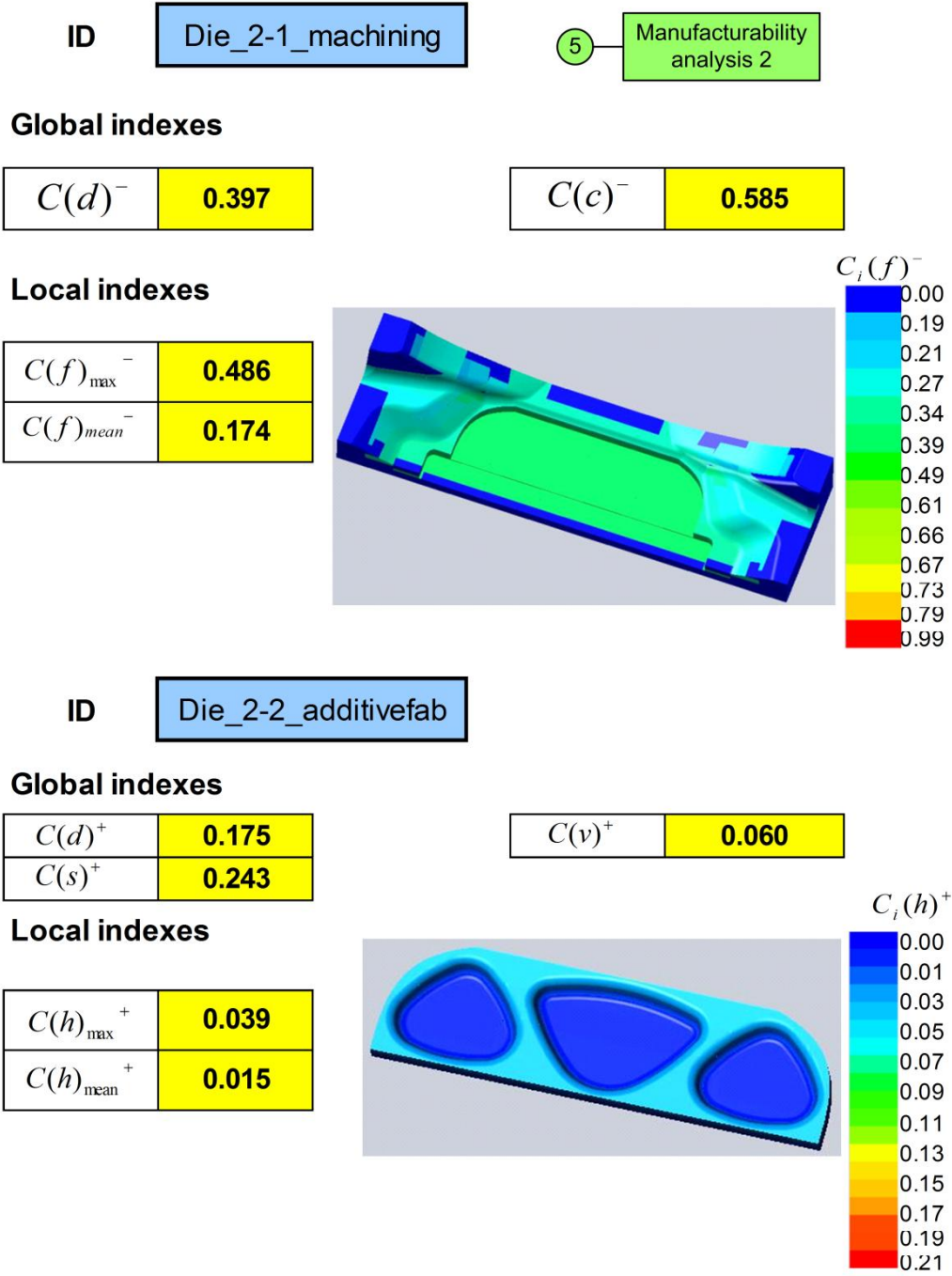

**Fig. 16.** Manufacturability analyses for the hybrid modular CAD model.

4.2.4. Comparison between one-piece design and hybrid one

Because the indexes are not equivalent (indexes for machining are not the same than indexes for additive manufacturing), no total values can be calculated. The comparison is done between the one-piece tool manufacturability analyses and the analyses for each module; it is presented in Fig. 17.

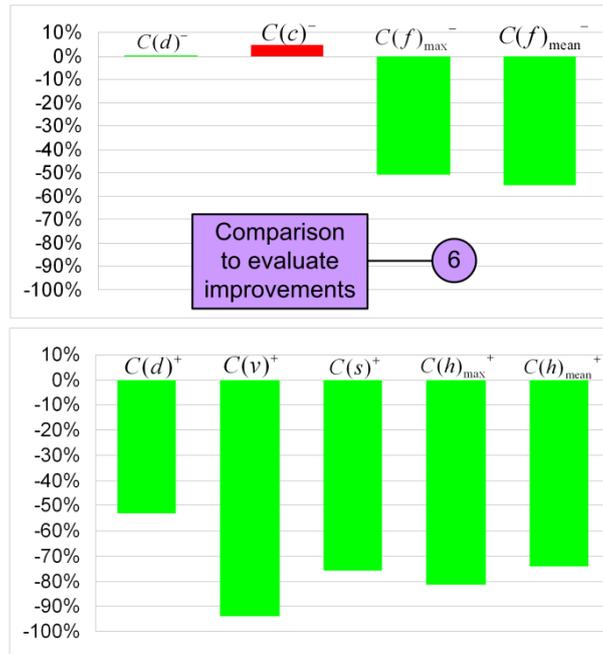

**Fig. 17.** Comparison between one-piece tool design and hybrid design.

The hybrid point of view has allowed designing a hybrid modular die in which manufacturability is improved. In fact, the map of manufacturing difficulties for module 1 shows that this module is easier to machine than the one-piece die. And the realization of module 2 by additive manufacturing process does not provide new manufacturing difficulties with regard to the manufacturability analysis of the one-piece die.

**5. Conclusions and objectives for further researches**

This paper presents a new hybrid modular design methodology. Starting from one-piece CAD model, global and local manufacturability indexes are calculated. In the case of local indexes, which provide a well-detailed view of which areas of the part are the most difficult to manufacture, the manufacturability analysis is based on octree decomposition. This new approach enables us to focus on the areas of the part that are the most difficult to manufacture because an accurate view of the manufacturing difficulties distribution is obtained. Then, hybrid and modular points of view help designers to choose between a one-piece design and a hybrid modular one.
This method has been developed as a new DFM system in CAD software. It is one of the first attempts to expand the DFM concept to a multi-process situation, combining additive processes (such as SLS or powder projection) to more traditional subtractive ones (HSM) in a hybrid modular vision. Two industrial examples taken from the field of tooling have been treated to illustrate the possibilities of this new methodology, and the way it can be used in an industrial manner.

Further research will be conducted to optimize the methodology and to define new manufacturability indexes. In this paper, indexes are based on geometric parameters. It is important to bring into play indexes that can be calculated directly at the design stage. Parameters that require a complete manufacturing preparation analysis (for example: cutting tool path strategy) are not taken into account, so as not to depend on manufacturer's skills. To have a more detailed view of manufacturing complexity, more accurate manufacturability indexes may be calculated, with other parameters involved, based on material information and technical specifications. A study has to be carried out in order to be able to compare different indexes between each other. The way the assembly constraints modify the design of the modules has also to be integrated in the methodology because all the modules must be carefully gathered in order to create a whole part with same level of quality as a one-piece part.